# A validation strategy for *in silico* generated aptamers

R. Cataldo[1,*], F. Ciriaco[2,*], E. Alfinito[3,*]

[1]Department of Mathematics and Physics,"Ennio de Giorgi", University of Salento, Via Monteroni, Lecce, Italy, I-7310,

[2]Department of Chemistry, University of Bari, Via Orabona 4, Bari, Italy, I-70126,

[3]Department of Innovation Engineering, University of Salento, Via Monteroni, Lecce, Italy, I-73100

*To whom correspondence should be addressed.

**Abstract** The selection of high-affinity aptamers is of paramount interest for clinical and technological applications. A novel strategy is proposed to validate the reliability of the 3D structures of aptamers, produced *in silico* by using free software. The procedure consists of three steps: a. the production of a large set of conformations for each candidate aptamer; b. the rigid docking upon the receptor; c. the topological and electrical characterization of the products. Steps a. and b. allow a global binding score of the ligand-receptor complexes based on the distribution of the "effective affinity", i.e. the sum of the conformational and the docking energy. Step c. employs a complex network approach (Proteotronics) to characterize the electrical properties of the aptamers and the ligand-receptor complexes. The test-bed is represented by a group of anti- Angiopoietin-2 aptamers. In a previous literature these aptamers were processed both *in vitro* and *in silico*, by using an approach different from that here presented, and finally tested with a SPS experiment. Computational expectations and experimental outcomes did not agree, while our results show a good agreement with the known measurements. The devised procedure is not aptamer-specific and, integrating structure production with structure selection, candidates itself as a quite complete theoretical approach for aptamer selection.

**Contact:** rosella.cataldo@unisalento.it, fulvio.ciriaco@uniba.it, eleonora.alfinito@unisalento.it

**Highlights**

- The 3D structures of 5 different anti-Angiopoietin aptamers are produced *in silico*
- The 3D structures are ranked by using a new indicator called "effective affinity"
- The affinity of an aptamer for its target is monitored by using a complex network
- The resistance of the aptamer-protein complex gives insights about affinity

# 1 Introduction

The growing interest in therapeutic aptamers (Lee *et al*., 2015) is driving research towards even more efficient and stable macromolecules. The biotechnological approaches, mainly the SELEX procedure, (Tuerk et al., 1990) involve high costs both in materials and time. Hard is also the problem of resolving the 3D structures: X-ray crystallography and NMR spectroscopy provide few and sometimes controversial data about aptamers (Sun and Zu, 2015). Furthermore, the crystalline state of the aptamers-protein complex could not accurately reproduce the shape assumed in solution (Li and Lu, 2009).

A large number of computational methods and applications (Chushak et al.,2009; Bini et al., 2011) have been developed, starting from the experience gained in predicting protein sequences and structures (Gilson et al., 2007; Rother et al., 2011). Due to the wide range of size and behaviour of ligands, aptamers as a special case, and of receptors, these methods derive from entirely different concepts and obtain different accuracy (Kitchen et al., 2004).

The affinity of an aptamer for a receptor depends on the reciprocal capability to attain geometrical conformations where the binding functionalities match each other (Kitchen et al., 2004). Methods are present in the literature (Kinnings et al., 2011) that completely avoid the geometrical docking problem, relying instead on classification of the ligands on the basis of a large number of molecular descriptors (Stewart and McCammon, 2006). However, even when a good set of descriptors can be found, the ligand classification must be benchmarked against a large number of known samples.

Renouncing to follow the equations of motion to concentrate only on the recognition of the lowest energy conformations of the ligand-receptor system, docking methods are important representatives of these approaches. In a nutshell, docking generates samples of the conformational space of the system and ranks them. Therefore, both the exhaustiveness of the sampling and the correctness of the ranking function ultimately affect the accuracy of docking (Kitchen et al., 2004).

Hu et al. (2015) computationally selected RNA mutant sequences with high affinity for Angiopoietin-2 (Ang2), starting from the sequences of anti-Ang2 aptamers, obtained by the SELEX procedure. Using the ZDOCK program, the Authors of (Hu et al., 2015) carried out simulations of aptamer-protein interactions, scoring the result of each simulation with the ZRANK functions in Discovery Studio 3.5 (DS 3.5; Accelrys Inc., San Diego, USA). To test the prediction accuracy, they performed measurements with a surface plasmon resonance (SPR) biosensor. The three highest ZRANK score mutant sequences along with a high (Seq1) and low (Seq16) affinity binding sequence were analysed. Quite interestingly, one of the mutant sequences, named Seq2_12_35, which showed the best ZRANK score among the five-selected aptamers, was one of the worst performing in experiments.

These outcomes highlight the challenge of the *in silico* determination of the 3D conformation of RNA aptamers and aptamer-protein complexes. This is significantly more difficult than protein structure determination (Doudna, 2000), so that the majority of known RNAs remain structurally uncharacterized (Boniecki et al., 2016).

Boniecki et al. (2016) developed a free software, SimRNA, for computational RNA 3D structure prediction. SimRNA uses a coarse-grained representation of the RNA skeleton. It then relies on Monte Carlo methods for sampling the configurational space, guided by a suitable potential energy, statistically derived from experimental data. For modelling complex 3D structures, the software can use additional restraints and constraints, derived from experimental or computational analyses, including information about secondary structure and/or long-range contacts. SimRNA can be also used to analyse conformational landscapes and identify potential alternative structures.

The modelling of the physical properties of biomolecules, that is, electrical transport, conformational change, thermal modes and so on, is a long time debated problem (Tirion, 1996; Baranowski, 2006; Piazza et al., 2009), mainly concerned with the level of granularity used for the description. Recently, a novel approach called *Proteotronics*, able to conjugate structure and function of proteins and aptamers at a microscopic level, has been developed (Alfinito *et al.,* 2009, 2011, 2015, 2017). The core idea is that structure and function of biomolecules can be described simultaneously, by using a complex network whose degree of connections depends on the biomolecule activation state (Alfinito et al., 2009, 2011, 2015, 2017).

Proteotronics, initially developed for proteins, was for the first time tested on the single DNA 15-mer thrombin binding aptamer (TBA) alone and complexed with thrombin. It correctly described and interpreted some relevant results obtained by experiments. In particular, the model was able to foresee the reduced affinity of the TBA-thrombin complex, when produced in the presence of $Na^+$, with respect to that of the same compound, produced in a solution containing $K^+$. Furthermore, the model revealed that resistance measurements are sensitive to different affinities (Alfinito et al., 2017).

This paper proposes a novel computational strategy for the screening of a group of aptamers, attempting an evaluation of their binding affinity for a receptor. This strategy is described and benchmarked in the following points:

- Sampling RNA-aptamer conformations (pre-docking) through an *ad hoc* computational procedure.
- Docking all the previously obtained aptamer sample conformations to the target.
- Capturing some topological and electrical features of the aptamer docked with the

target, by using the principles of *Proteotronics*.

• Comparing the theoretical results with experiments (Hu *et al*., 2015).

The entire procedure shows a satisfactory agreement with the experimental findings, so that it can be considered successful when used for *in silico* aptamer docked structures validation.

## 2 System and Methods

The method here proposed was applied to the same problem as in (Hu et al., 2015), that is a comparative evaluation of binding to Ang2 of five different aptamers:

1. an aptamer, denoted "Seq1", both in Hu's paper and below, from the pool of Ang2 specific RNA aptamers known in the literature;

2. three mutant sequences, here and in Hu's paper denoted "Seq2_12_35", "Seq15_12_35", and "Seq15_15_38";

3. an Ang1-specific RNA aptamer, denoted "Seq16", as in Hu et al., there applied as a control sample.

### 2.1 *Sampling RNA-aptamer conformations*

Among the tools for the prediction of RNA tertiary structure (Dawson et al., 2016), SimRNA (Boniecki et al., 2016) was chosen.

SimRNA makes use of a simplified (coarse-grained) representation of the nucleotide chain, consisting of 5-6 dihedral angles for each nucleotide to describe the general aspects of the chain. The program then applies a Monte Carlo scheme for sampling the conformational space, with acceptance and rejection, dictated by a function that plays the role of potential energy.

The function prescription is obtained from a large set of crystallographic well resolved structures (Boniecki et al., 2016).

For each prediction, after an initial annealing phase, we carried out four independent runs of the Replica Exchange Monte Carlo simulation (Boniecki et al., 2016), each employing ten replicas.

Then, we performed a clustering of the obtained structures, based on geometrical similitude, following the procedure drawn in the SimRNA manual (Boniecki et al., 2014).

By means of the clustering algorithm, from about 30 up to 60 clusters were produced for each studied aptamer. A finer sampling can be obtained by tweaking the parameters. The structures were statistically analysed by proper SimRNA functions, in order to obtain, for each of the five studied aptamers, the frame corresponding to the lowest energy (Boniecki et al., 2014).

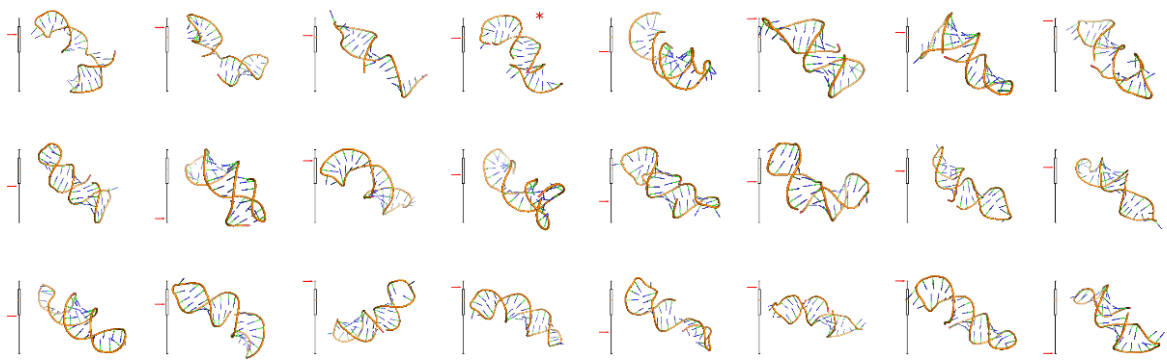

**Figure 1.** *A catalogue of Seq1 conformations, each representative of a different cluster, as obtained from SimRNA. On the left of each cartoon an annotation of its SimRNA energy. The picture marked with * is an example of configuration for which SimRNA does not provide a back-mapping sufficiently accurate to allow PyMol and MGLTools to recognize the structure as a single sequence.*

In doing so, the code permits to establish a lower RMSD threshold, for the first pass of clustering, and a higher RMSD threshold for a second pass of clustering (Boniecki et al., 2014). Usually, those thresholds are about 10% of the number of nucleobases of the sequence.

The receptor binding domain 1Z3S of Ang2 (Barton et al., 2005) and the aptamers were assigned partial charges and atom types by means of MGLTools.

The backmapping of the SimRNA reduced set of freedom degrees to the full set of atom coordinates was in a few cases unsuccessful: MGLTools in particular did not recognize the reconstructed molecule as a single sequence, due to infringement of geometrical constraints on bond distances. For example, the structure marked * in Figure 1 is broken into two subsequences, both by MGLTools and by PyMol, as visible in those screenshots.

We choose to discard such structures rather than repairing them with *ad hoc* procedures or modifying the binding parameter of MGLTools. MGLTools was also used to translate back the poses obtained from AutoDock-Vina to pdb format.

Each of the aptamer conformations obtained from SimRNA and validated by MGLTools was therefore rigidly docked to the Ang2 receptor. The docking of the aptamers to the receptor was performed by means of AutoDock-Vina (Trott et al., 2010).

*2.2 The Proteotronics approach*

The Proteotronics approach is a theoretical and computational procedure to analyse the physical response of biomaterials in electronic devices. It is a single-macromolecule modelling founded on the *structure and function paradigm*, born to describe the macroscopic data as emerging properties due to local interactions. The general strategy rises to the macroscopic physical features, by using a coarse-grained description of the 3D structure. In the literature (Tirion, 1996; Baranowski, 2006; Piazza et al., 2009; Alfinito et al., 2015), the level of refinement of this kind of description ranges from the complete molecule to the single atom. A good compromise is observed in the case of the single amino-acid level, sufficient to keep most of information useful for technological applications, with the advantage of quite small computational time. This kind of description has been extended to aptamers (Alfinito et al., 2017), since the macroscopic mechanism of aptamer binding is quite similar to that observed in proteins.

The procedure has been extensively described in previous papers (Alfinito et al., 2009, 2011, 2015, 2017) and consists of three steps:

- The graph analogue building;

- The interaction network building;

- The network solution.

Starting from the 3D structure of the aptamer, the corresponding graph is set up, using the following rules:

1. Each nucleobase (amino acid) is mapped into a single node, whose space position is that of $C_1$ ($C_\alpha$) carbon atom taken as the centroid of the real molecule (Alfinito et al., 2015). Two nodes are connected with a link only if their distance is below an assigned interaction radius, $R_C$. The graph preserves the macromolecule topology. A sketch of one of the possible representations of Seq 1 and the corresponding network, calculated with $R_C=20$Å, is reported in Figure 2.

2. A specific kind of interaction is selected and associated to each link. Here, a simple charge transfer in the linear regime is described. Each link mimics an electrical pathway with a specific elementary resistance. The resistance between a couple of nodes, say $a,b$, is calculated as that of a cylindrical structure of length $l_{a,b}$, the distance between the nodes, and surface $A_{ab}$, the intersection area of the spheres of radius $R_C$, drawn around the nodes.

3. Finally, resistance can be calculated by using appropriate resistivity values, as detailed in (Alfinito et al., 2017). A couple of ideal electrodes connects the network to an external bias. The network is solved, for an assigned value of $R_C$, by using the standard Kirchhoff rules.

## 3 Results and Discussion

### 3.1 Effective affinity

For the following discussion, we introduce the term "effective affinity" (EA) to indicate

$$EA = E_{docking} + E_{SimRNA}. \tag{1}$$

A justification is here needed on how we misuse the SimRNA knowledge-derived potential to translate it into energetic units and how it contributes to the global ranking.

The SimRNA potential is the guide of Monte Carlo procedure, driving the structure from the initial state to the most stable conformation (Boniecki et al., 2016). Therefore, it effectively works by ranking the conformations on the basis of their energy, though a large imprecision has to be expected and perfectly in line with the purpose of the SimRNA potential. The function structure is obtained from scrutiny of a large number of experimental RNA structures, so as to best match the distribution of local configurational motifs of SimRNA *in silico* evolution and the experimental distribution of the same motifs in the selected database.

Therefore, we assimilate the SimRNA potential function to a sort of approximated thermodynamic potential, accountable for the statistical distribution of conformational parameters.

Since most of the experimental structures behind the SimRNA potential are reasonably obtained at room temperature, we decided to translate the unitless SimRNA energy on the basis of the formula:

$$E_{SimRNA} = E_{unitless} \cdot \mathcal{R}T \qquad (2)$$

where $\mathcal{R}$ is the gas constant and T=298 K.

Among the thermodynamic functions, the effective affinity should have the closest correspondence with the binding enthalpy.

As approximate as it may be, the SimRNA energy contribution cannot be discarded in the evaluation of ligand affinity, unless one finds a better evaluation of the aptamer conformational energy.

The distribution of docking energies in Figure 4 is meant to illustrate this concept: docking energies much lower than those corresponding to the most stable aptamer conformation are present, others could appear if the sampling procedure were extended so that higher energy conformations are represented.

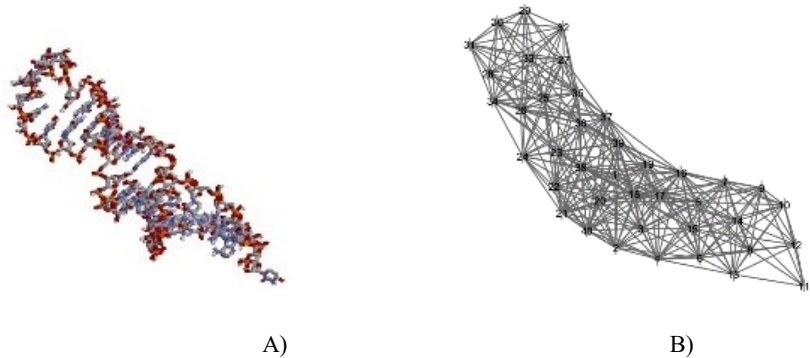

A)                                          B)

**Figure 2**. *A) The aptamer Seq1 in one of its conformations and B) the corresponding graph ($R_C$=20Å).*

A more common strategy is to obtain the most stable ligand conformation somehow as a starting point for docking computation. However, a benchmark conducted on NPDock, a web server specialized in protein-aptamer docking (Tuszynska et al., 2015), with a large number of protein-RNA complexes, showed that only about one half of the docked structures reproduced the native ones, for the easy targets, whereas the matches of experimental and *in silico* structures where negligible, for the difficult targets.

The meaning of the present approach is that, by starting from a large sample of ligand conformations, it is possible to obtain more stable docked structures than proceeding from the minimum energy conformation. The slideshow in Figure 1 shows a set of Seq1 conformations, all reasonably affordable at room temperature.

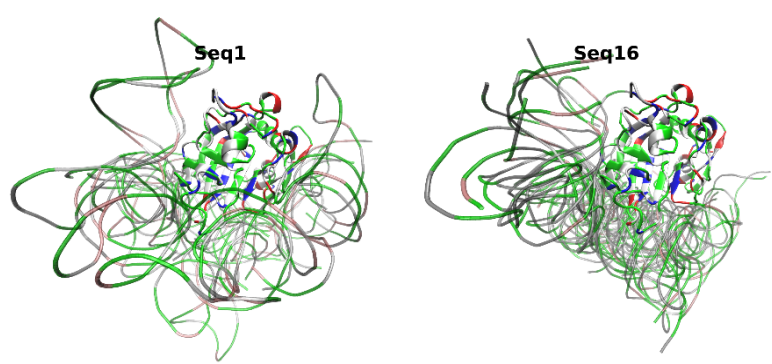

**Figure 3**. *Overlapped best docking positions to Ang2 for the set of conformations of Seq1 and Seq16, generated by SimRNA.*

Figure 3 shows the behavior of a subset of Seq1 and Seq16 conformations, docked to Ang2. It is evident that aptamers with different conformations have different preferred docking positions, though an important crowding around a few specific spots is present for Seq1. Less selectivity is displayed by Seq16 instead, resulting in a much more uniform crown of docking positions.

In Figure 4, the red arrows represent the effective affinity in correspondence of the minimum energy aptamer conformation. Since we referred the aptamer energy conformation to its minimum, on red arrows the effective affinity equals the docking energy. For sequences Seq1 and Seq2_12_35, the red arrows are significantly displaced from the minimum effective affinity, which is therefore obtained from a different ligand conformation.

Black arrows represent the ZRANK results as reported by (Hu et al. 2015). Red arrows and black arrows represent entirely equivalent concepts, the difference in values being due to different choices of the computational software, both to obtain aptamer conformations and to dock them to the receptor, probably mainly in the effective docking potential.

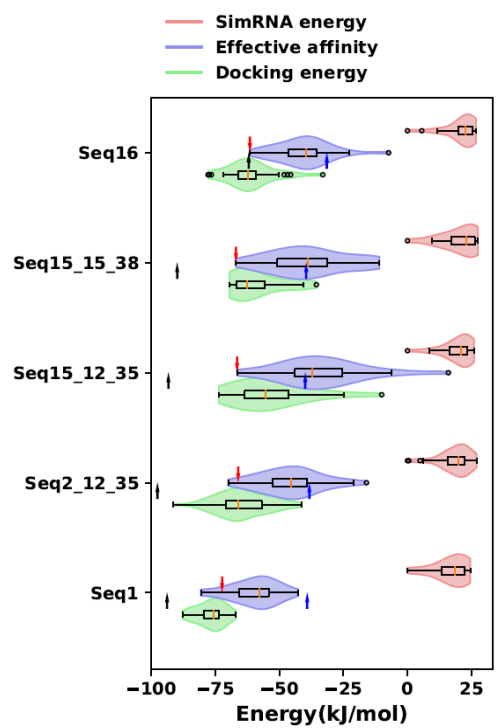

*Figure 4*. Box plots with outliers and kernel density plots for various computed energy distributions: docking energy (green), SimRNA energy (pink) and effective affinity (violet). Blue arrows represent the experimental $\Delta G^0$, black arrows the ZRANK values from Hu et al. (2015), red arrows the docking energy/effective affinity in correspondence of the most stable aptamer conformation.

Unfortunately, such differences should be expected in the present state of the docking art, as equally to be expected are important deviances from experimental data: blue arrows point to the standard free energies $\Delta G^0$ obtained from the binding constants reported in (Hu et al. 2015). More important, however, is the possibility to obtain a similar trend for experimental and simulated binding energies.

An inspection of the experimental binding data shows that sequences Seq1, Seq2_12_35, Seq15_12_35 and Seq15_15_38 behave similarly. The differences among their $\Delta G^0$, are too small to be reliably reproduced in docking calculations, or by any other computational tool; they are also irrelevant to any practical application.

Seq16 instead displays weaker binding. This aspect is well reproduced both in the present calculations and in those by Hu *et al.* (2015), though in the latter case the difference is more evident.

The present calculations however estimate a sensibly stronger binding for Seq1 than for all other sequences. This might well be coincidental, given the small number of aptamers considered, but we would like to advance also two possible causes:

1. wild aptamers could effectively have been engineered by natural selection to span a smaller configurational space;

2. the knowledge-based potential adopted by SimRNA, being obviously based on natural sequences, builds a better potential for wild aptamers, e.g. taking somehow better into account long range interactions (Boniecki et al., 2016).

The different aptamer conformations were then analysed with respect to their topological and electrical properties. These are powerful tools to identify the mean characteristics of a sequence and to detect extreme structures. Finally, they can be used to make a comparison among sequences.

*3.2 Topological properties*

To explore the backbone topology of the sequences, we can refer to the contact map, i.e. the graphical representation of the adjacency matrix (Albert et al., 2002). There are not significant differences among the structures corresponding to the same sequence, both in the pre-docking and the post-docking phase. In particular, the two possible choices of the RMSD threshold (Section 3.1) produce quite similar results.

A selection of contact maps, one for each sequence, is reported in Figure 5, for $R_C$=20 Å. They represent the structures of the protein-free aptamer in the post-docking phase. A qualitative similarity of all the sequences but Seq1 can be argued. Seq1 shows two branches corresponding to the double twist, already shown in Figure 1, while the single branch of the other sequences describes a simple hairpin. In conclusion, the 5 sequences fold in a quite different way, and this is one of the elements to be considered in the evaluation of the affinity for Ang2. Specifically, a different folding exposes a different surface useful for binding.

The docked structures were also analysed. In particular, some global information about topology is given by the number of links of the aptamer-protein-analogue network: the larger the number of links, the closer the aptamer is to the protein. This gives an estimate of the protein-aptamer complementary, although not of the stability of the binding.

The link number was calculated for different $R_C$ values. The Spearman rank correlation can be used to evaluate the results for different values of $R_C$. It shows a strong, and in some cases very strong correlation between the link number and the docking energy /effective affinity (See Table 1), thus confirming that both these quantities give a good estimation of the protein-aptamer complementary in structure. The best result is given by $R_C$ =20 Å. Figure 6 reports the corresponding data.

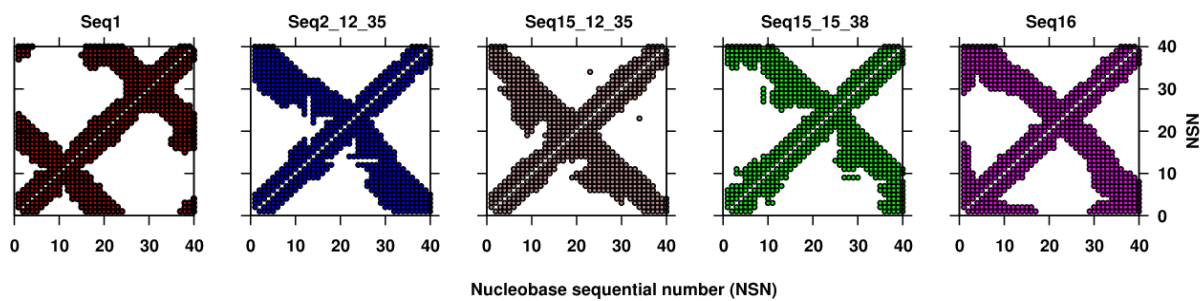

**Figure 5**. *Contact maps for the studied sequences ($R_C$=20Å).*

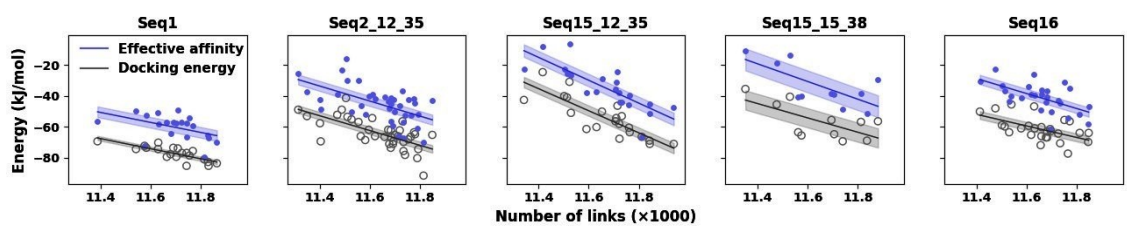

**Figure 6**. *The total number of links vs. docking energy and effective affinity. The interaction radius is $R_C$=20 Å.*

*3.3 Electrical properties*

The resistance spectrum has been calculated for each structure, over an assigned range of $R_C$ values (Alfinito *et al.* 2009, 2011, 2015, 2017). By increasing $R_C$, the network link number grows and the resistance decreases. The resistance of the protein-free aptamer in the post-docking phase strongly depends on the shape of the aptamer. Therefore, it could be considered a measure of the surface that the aptamer effectively offers to the protein (effective surface). In fact, in a simple circuit analogue, a surface, ***S***, can be associated to each resistance, ***R***, so that, referring for the sake of simplicity to a cylindrical geometry, ***S*** ~1/***R***.

| Sequence | Rank | Significance | Rank | Significance |
|---|---|---|---|---|
| | Docking energy, $R_C=10$Å | | Docking energy, $R_C=20$Å | |
| 1 | -0.37 | 1.1e-1 | -0.78 | 7.1e-4 |
| 2_12_35 | -0.58 | 1.5e-4 | -0.63 | 1.3e-4 |
| 15_12_35 | -0.56 | 1.1e-2 | -0.71 | 1.2e-3 |
| 15_15_38 | -0.61 | 5.4e-2 | -0.61 | 5.4e-2 |
| 16 | -0.43 | 2.9e-2 | -0.51 | 9.5e-3 |

| Sequence | Rank | Significance | Rank | Significance |
|---|---|---|---|---|
| | EA, $R_C=10$Å | | EA, $R_C=20$Å | |
| 1 | -0.55 | 1.7e-2 | -0.53 | 2.1e-2 |
| 2_12_35 | -0.50 | 2.5e-3 | -0.57 | 4.8e-4 |
| 15_12_35 | -0.71 | 1.6e-3 | -0.89 | 6.4e-5 |
| 15_15_38 | -0.79 | 1.3e-2 | -0.52 | 1.0e-1 |
| 16 | -0.46 | 1.8e-2 | -0.57 | 3.6e-3 |

**Table 1**: *Spearman correlation between the docking energy and the effective affinity, for $R_C =10$ Å and 20Å.*

In Figure 7A (on top), a bar plot reports the mean resistance of each protein-free sequence in the post-docking phase calculated at $R_C=20$Å, and given in arbitrary units, i.e. normalized to the largest value (Seq16). Seq1 and Seq15_15_38 show the lowest resistance, i.e. the largest effective surface, and Seq16 exhibits the largest resistance instead. In Figure 7A (on bottom) the ratio of the aptamer-protein resistance to the aptamer resistance is also reported: in this case, the complex Seq16-Ang2 has the lowest resistance ratio, while Seq15_12_35 and Seq15_15_38 the highest. This result can be interpreted looking at the analogue electrical circuit: specifically, the protein is represented by a resistor ladder with the number of ladders increasing with $R_C$, the aptamer-protein complex is represented by the parallel circuit of the protein and the aptamer resistance. The resistance of the complex, $R_{comp}$, is the equivalent resistance of the parallel circuit, smaller than both aptamer ($R_{apt}$) and protein resistance ($R_{prot}$), and the corresponding effective surface is larger than the aptamer, $S_{apt}$, and the protein, $S_{prot}$, effective surface. The cartoon of these analogue circuits is reported in Figure 7B. Finally, the ratio $R_{comp}/R_{apt} = (1+(R_{apt}/R_{prot}))^{-1}$ can be interpreted as the percentage of contact surface. As a matter of fact, it is always smaller than 1, larger than 0.5 only if $R_{apt}$ is smaller than $R_{prot}$, i.e. $S_{apt}$ is larger than $S_{prot}$ and the protein can be hosted in the aptamer binding site. In case $R_{comp}/R_{apt}$ is smaller than 0.5 the binding should not happen ($S_{prot}$ is larger than $S_{apt}$ and the protein cannot be hosted). In the studied case, high affinity sequences have a large value of the ratio $R_{comp}/R_{apt}$. A cartoon about these correspondences is reported in Figure 8.

Finally, looking at Figure 7A, we conclude that the highest affinity has to be attributed to Seq15_12_35 and the lowest to Seq16, in agreement with the results presented in the literature (Hu et al., 2015). However, it has to be highlighted that Seq15_15_38 and Seq1 show a response quite close to the best one.

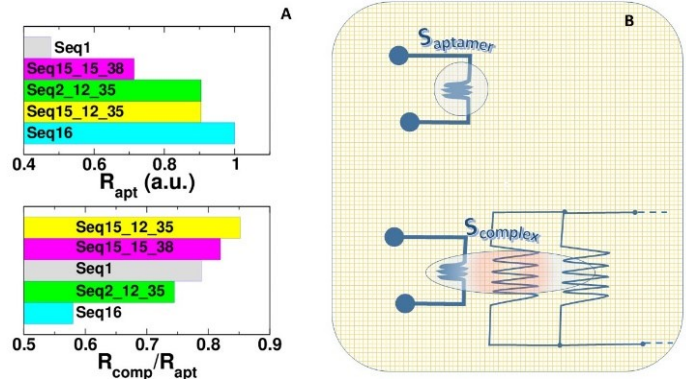

**Figure 7.** *Mean resistance of 5 selected sequences. A. On top: the resistance of the protein-free aptamer in the post-docking phase; on bottom: the ratio of the complex to the resistance of the protein-free aptamer. B. Cartoon of the corresponding circuits and the associated effective surfaces $S_{apt}$ of the aptamer. and $S_{comp}$ of the complex ($R_C$=20Å).*

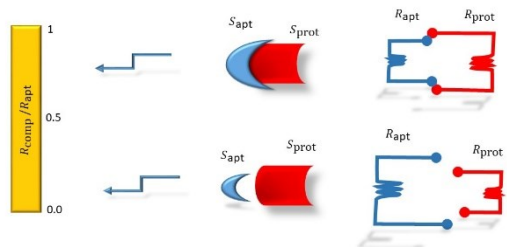

**Figure 8.** *Cartoon of two possible aptamer-protein interfaces, with the corresponding circuits.*

*3.4 PCA analysis on the docked structures*

As an alternate strategy for ranking the structures produced *in silico*, we performed some other statistical analyses to provide several possible markers, very fast to calculate. In doing so, the Principal Component Analysis (PCA) was employed, a powerful and common technique for finding patterns in high dimensional data, extensively applied in fields such as face recognition and image compression (Gonzalez and Woods, 2017).

The main advantage is that, once these patterns in the data are found, especially when we have high dimensional samples, it is possible to compress the input, i.e. to reduce the number of dimensions, with a modest loss of information in describing the whole system.

In this context, PCA was applied in order to:

- isolate, within a specific sequence, those structures that contain the major amount of information (characterization);
- identify those structures that are quite similar from a statistic point of view;
- determine those structures in which electrical features (resistances) have high correlation with docking energies.

For the considered five sequences, about 600 structures were obtained after the docking phase. For each structure, a resistance calculation for 100 different $R_C$ values, ranging from 10 to 110 Å was performed.

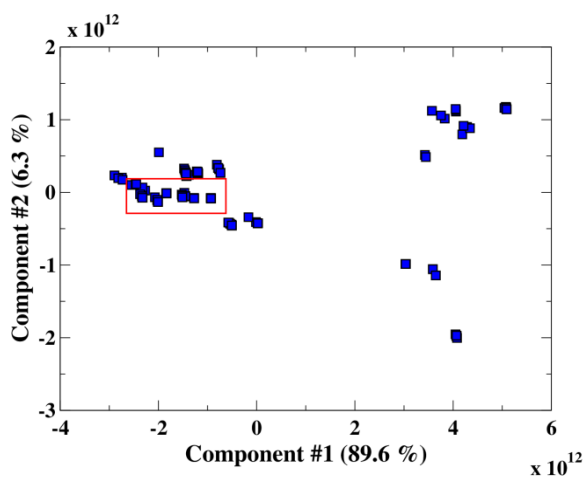

**Figure 8**. *PCA results for the Seq15_12_35. The box indicates the closest representations.*

Therefore, it is possible to construct a vector of features comprising the docking energy, called affinity in the AutoDock-Vina log file, together with the RMSD for the lowest and upper bound, the resistivity values, obtained for both the ligand and the ligand-receptor complex, the difference and the relative difference of those resistivities.

For each considered sequence the first component explained over 70% of information, the structures could be well differentiated; resistances seem to be well correlated with docking energy. Therefore, a shortlist of the closest structures, in which the electrical features have high correlation with docking energies, can be drawn.

An example is given in Figure 8, where it is evident that the structures representing Seq15_12_35 follow in three macro-areas, with different distances among them. Specifically, the closest structures are highlighted with a box in the same figure. This is a quite useful and powerful tool for evaluating the *corpus* of structures of a single aptamer, because it takes into account all the calculated information.

Furthermore, we assessed that the structures within the same cluster were quite similar, and, for $R_C$ values greater than 20 Å, the principal components do not significantly change, accordingly with the observations highlighted previously.

The method is reasonably fast. For example, on a Xeon 6-Core E5-2620v2 2.1Ghz 16 MB of RAM, the SimRNA simulation time reported in the log-file is about 12 hours, for sequences of 41 nucleotides. As regards the docking phase, AutoDock-Vina employs about 36 cpu*h for each structure. Considering that each SimRNA or AutoDock-Vina run is independent, the procedure can be easily automated to screen out a large number of aptamer sequences/structures. The Proteotronics computational time is of few minutes for each structure.

## 4. Conclusions and discussions

In conclusion, in this paper we have used a recent method (Boniecki et al., 2016) for the *in silico* generation of the 3D structures of a set of 5 anti-angiopoietin aptamers, specifically, Seq1, a known anti-Ang2 aptamer, Seq16, a known anti-Ang1 aptamer and 3 mutant structures, Seq15_15_38, Seq15_12_35 and Seq2_12_35. The aptamer linear sequences were given in (Hu et al., 2015) and the docking with Ang2 was performed by using a set of rigid rotations (Trott et al., 2010).

A statistic investigation of the results was performed by using several techniques to identify indicators useful to assess the aptamer affinity for Ang2. An electrical network analogue of the aptamer and the protein was set up, able to explore their topological properties.

A novel energy-like quantity, called *effective affinity* is proposed as an appropriate indicator of the aptamer affinity for Ang2. The high correlation with a topological indicator like the network link number, which measures the closeness of the aptamer-protein complex, confirms this proposition. The link number gives only a global information about the structure, therefore, to estimate the space distribution of links, i.e. to have a local information about structure, the resistance of the electrical network analogue in the linear regime is calculated. Seq 1 shows the most complex structure as backbone, and has also the lowest resistance, while the resistances of all the other structures are comparable. On the other hand, looking at the complexes, we can note a strong difference in Seq16 which has the lowest ratio of the complex to the aptamer resistance. This result has been interpreted in terms of the percentage of contact surface, which is quite large in the anti-Ang2 specific aptamers and small in the non-specific anti Ang2 aptamer, Seq 16. These results confirm resistance as a good tool for investigating chemical affinity.

Finally, the PCA technique allows us to select structures which have a similar behavior and which can be used to represent the real aptamer.

The devised computational procedure is not aptamer-specific, and has the major improvement, with respect previous investigation, of integrating different theoretical techniques. Most importantly, the ranking provided by the present procedure is in reasonable agreement with experimental data.

**Acknowledgments**

We thank the financing project Laboratorio SISTEMA, "Laboratorio per lo Sviluppo Integrato delle Scienze e delle Tecnologie dei Materiali Avanzati e per dispositivi innovativi" PONa3_00369 for providing the computational resources for the docking calculations.